\documentclass[fleqn,usenatbib]{mnras}

\usepackage{newtxtext,newtxmath}

\usepackage[T1]{fontenc}

\DeclareRobustCommand{\VAN}[3]{#2}
\let\VANthebibliography\thebibliography
\def\thebibliography{\DeclareRobustCommand{\VAN}[3]{##3}\VANthebibliography}

\usepackage{graphicx}	
\usepackage{amsmath}	
\usepackage{amssymb}	

\title[OH absorption with FAST]{A pilot search for extragalactic OH absorption with FAST}

\author[Z. Zheng et al.]{
Zheng Zheng$^{1}$\thanks{E-mail: zz@bao.ac.cn (ZZ)},
Di Li$^{1\,2\,3}$\thanks{E-mail: dili@nao.cas.cn},
Elaine M. Sadler$^{4,5,7}$,
James R. Allison$^{6,7}$
and Ningyu Tang$^{1}$
\\
$^1$ National Astronomical Observatories, Chinese Academy of Sciences, 20A Datun Road, Chaoyang District, Beijing 100101, China \\ 
$^2$ University of Chinese Academy of Sciences, Beijing 100049, China\\
$^3$ NAOC-UKZN Computational Astrophysics Centre, University of KwaZulu-Natal, Durban 4000, South Africa \\
$^4$Sydney Institute for Astronomy, School of Physics, University of Sydney, NSW 2006, Australia\\
$^5$ATNF, CSIRO Astronomy and Space Science,  PO Box 76, Epping, NSW 1710, Australia \\
$^6$ Department of Physics, University of Oxford, Denys Wilkinson Building, Keble Rd., Oxford OX1 3RH, UK\\
$^7$ARC Centre of Excellence for All Sky Astrophysics in 3 Dimensions (ASTRO 3D) 
}

\date{Accepted XXX. Received YYY; in original form ZZZ}

\pubyear{2020}

\begin{document}
\label{firstpage}
\pagerange{\pageref{firstpage}--\pageref{lastpage}}
\maketitle

\begin{abstract}
OH absorption is currently the only viable way to detect OH molecules in non-masing galaxies at cosmological distances. There have been only 6 such detections at $z>0.05$ to date and so it is hard to put a statistically robust constraint on OH column densities in distant galaxies. We carried out a pilot OH absorption survey towards 8 associated and 1 intervening HI 21-cm absorbers using the Five-hundred-meter Aperture Spherical radio Telescope (FAST). We were able to constrain the OH abundance relative to HI ([OH]/[HI]) to be lower than $10^{-6}\sim 10^{-8}$ for redshifts  z$\in$ [0.1919, 0.2241]. Although no individual detection was made, stacking three associated absorbers free of RFI provides a sensitive OH column density 3-$\sigma$ upper-limit $\sim 1.57 \times 10^{14} (T_x^{\rm OH}/10K)(1/f_c^{\rm OH})cm^{-2}$, which corresponds to a [OH]/[HI] < $5.45\times 10^{-8}$. Combining with archival data, we show that associated absorbers have a slightly lower OH abundance than intervening absorbers. Our results are consistent with a trend of decreasing OH abundance with decreasing redshift. 
\end{abstract}




\section{Introduction}
The hydroxyl radical (OH) is one of the most widely distributed molecules  in the interstellar medium (ISM) of our Galaxy \citep[e.g.][]{turner79, allen15,Li18b}. As an abundant simple hydride, OH was the first molecule to be detected in the radio band \citep{weinreb63} and is expected to trace the total amount of H$_2$ before all carbon is locked into CO \citep{vandishoeck88}.

Galactic studies \citep{Li18b,nguyen18} show that OH traces H$_2$ better than CO in a substantially bigger fraction of the ISM volume, but the normal thermolized OH need to be detected through absorption due to its excitation conditions. These results are consistent with the analysis of diffuse gamma-ray emission from Fermi \citep{Remy17}, which also revealed wide-spread molecular gas that cannot be traced by CO, thus dubbed `dark' molecular gas  \citep[DMG,  ][]{grenier05}.

Deep CO observations \citep[e.g.][and references therein]{Tacconi20} show that dense molecular content decreases since redshift $z\sim 1.4$.
Detections of OH lines in galaxies at cosmological distances can provide further constraints on both `dark' molecular gas fractions of these galaxies \citep[e.g.][]{Gupta18} and changes in fundamental constants at cosmological distances \citep[e.g.][]{Kanekar05,Kanekar12,Kanekar18,Curran04}. 
The inferred molecular gas fraction is also important for understanding star formation rate and galaxy evolution, e.g. why the star formation rate has decreased by a factor of $\sim10$ since $z=2$ \citep{MD14}. However, there have been only 6 OH absorbers at cosmological distances detected up to date. Four of these detections are intervening absorbers: B0218+357 \citep{Kanekar03}, G0248+430 \citep{Gupta18}, PKS 1830-211 \citep{Chengalur99,Allison17,Gupta20} and PMN J0134-0931 \citep{KB03,Kanekar05}. The other two are associated absorbers within quasars: B3 1504+377 and PKS 1413+135 \citep{KC02}.
Details about these detections are listed in Table 1.

\begin{table*}
	\caption{Previously detected OH absorbers at cosmological distances }
	\label{tab:archival_OH}
	\begin{center}
	\begin{tabular}{lccccccccccc} 

\hline
Source name & Type  & $z_{\rm abs}$   &  $\tau_{\rm HI,peak}$ & $\int\tau_{\rm HI}$ dv & HI FWHM & N(HI) & $\tau_{1667,\rm peak}$ & $\int\tau_{1667}$ dv & N(OH) & Refs \\
 &  &  & & (km.s$^{-1}$) & (km.s$^{-1}$) &  ($10^{20} \rm cm^{-2}$) 
 & ($10^{-2}$) & (km.s$^{-1}$) & ($10^{15} \rm cm^{-2}$) & & \\
& (1) & (2) & (3) & (4) & (5) & (6) & (7) & (8) & (9) & (10)  \\
\hline

PKS 1830-211 & I &  0.88582 & 0.055 & 5.8 & 96 & 31.72 & 0.7 & 1.83 & 11.39$^{\rm b}$ & C99,KC02 \\
B0218+357 & I &  0.68468 & 0.05 & 2.94 & 55 & 5.35 & 1 & 0.40 & 2.24 & K03,KB03,C03 \\
B3 1504+377 & A & 0.67343 & 0.32 & 21 & 62 & 51.73 & 0.83 & 0.448 & 2.18 & C97,KC02 \\
PKS 1413+135 & A & 0.24671 & 0.34 & 7.13 & 18 & 194.97 & 0.49 & 0.023 & 1.17 & C92,KC02\\
PMN J0134-0931 & I & 0.7645 & 0.047 & 7.06 & 142 & 25.74 & 1.8 & 2.19$^{\rm c}$ & 4.9 & KB03,K12 \\
G0248+430 & I   & 0.0519 & 0.02 & 0.43 & 20 & 0.55 & 0.4 & 0.08 & 0.063 & G18 \\

\hline
	\end{tabular}
	\end{center}
    \begin{flushleft}
      {\small 
Columns: 
(1) Absorber type, I for intervening and A for associated. 
(2) Redshift of the absorber. 
(3) HI peak optical depth, calculated using $1.06 \int\tau_{\rm HI} dv$ divided by the HI 21-cm absorption full-width-half-maximum (FWHM) \citep{Allison13} if not provided by the references. 
(4) Integrated HI optical depth. 
(5) HI 21-cm absorption FWHM, calculated using $1.06 \int\tau_{\rm HI} dv / \tau_{\rm HI,peak}$ \citep{Allison13} if not provided by the references. 
(6) HI column density. The HI excitation temperatures and covering factors used in the calculations are from the listed references. They are $T_{x}^{\rm HI} = \,$100, 100, 100, 150, 200, 70 K; $f_c^{\rm HI} = \,$0.33, 1, 0.74, 0.1, 1, 1, respectively. 
(7) OH 1667 MHz line peak optical depth. 
(8) Integrated OH 1667 MHz line optical depth. 
(9) OH column density. The OH excitation temperatures and covering factors used in the calculations are from the listed references. All OH excitation temperatures are 10K except for G2048+430, which has  $T_{x}^{\rm OH} = 3.5 \rm K$. The OH covering factors are $f_c^{\rm OH} = \,$0.36, 0.4, 0.46, 0.044, 1, 1, respectively. (10) References:C99: \citet{Chengalur99}; KC02: \citet{KC02};  K03: \citet{Kanekar03}; C03: \citet{Cohen03}; C97: \citet{Carilli97}; KC02: \citet{KC02}; C92: \citet{Carilli92}; KB03: \citet{KB03}; K12: \citet{Kanekar12}; G18: \citet{Gupta18}. \\

Notes: (a) Redshift of the background QSO is from  SIMBAD Astronomical Database \citep{Wenger00} and \citet{BH01}. 
(b) This value is from KC02 which used the parameters listed above. \citet{Gupta20} derived a lower OH column density, $1.46 \times 10^{15} \rm cm^{-2}$, using the data from the MeerKAT-64 array with a lower $T_{x}^{\rm OH} = 5.14 \rm K$ and a higher $f_c^{\rm OH} = 1$. 
(c) Estimated using the Gaussian fitting parameters listed in K12.
}
\end{flushleft}

\end{table*}

Most of these OH absorbers show an OH column density about the order of $10^{15} \rm cm^{-2}$.  \citet{Gupta18} show that galaxies with $z<0.4$ could have much lower OH column densities, i.e. below $10^{14} \rm cm^{-2}$, using a detected OH absorber and upper-limits of 8 non-detections and assuming the OH excitation temperature $T_x^{\rm OH} = 3.5\,$K, however, a more commonly used $T_x^{\rm OH}$ for OH absorbers at cosmological distances is 10 K \citep[e.g.][]{KC02,Kanekar03,Curran11}. 
The \citet{Gupta18} sample contains only intervening absorbers with relatively low redshifts. \citet{Grasha20} searched 11 intervening and 5 associated HI absorbers for OH absorptions and report no new detections except for a re-detection of the 1667 MHz OH absorption towards PKS 1830-211.
\citet{Grasha19} searched a large sample of compact radio sources for associated HI and OH absorptions but only 6 of them, which have relatively larger redshifts ($z\sim 0.6$), have HI absorption detections and none of them have OH absorption detection.
Intervening absorbers are mostly from foreground galaxies or HI dark clouds. Associated absorbers can be found in AGN outflows, circumnuclear disks, and cold gas clouds in the host galaxy or merger relics \citep{maccagni17,Oosterloo19,Allison19}. 
The cold molecular gas in these systems are not well studied due to a low detection number \citep{Allison19} and they could potentially have very different OH column densities and different excitation temperatures \citep{Curran16}. 
Therefore, we observe a sample of low-redshift ($z<0.3$) galaxies with associated HI 21-cm absorption detections, with an extra target with a deep intervening HI 21-cm absorption for comparison, using the Five-hundred-meter Aperature Spherical radio Telescope \citep[FAST;][]{Nan11,LP16,li18-fast,Jiang20} aiming to study their cold gas parameters through OH absorptions. 

The outline of the paper is as follows: we describe the sample and observation strategy in Section \ref{sec:sample}; we then present the data reduction procedure and standing wave removal algorithm in Section \ref{sec:data}; we show our results and discussions in Section \ref{sec:results} and summarize them in Section \ref{sec:summary}. An HI 21-cm absorption spectrum derived from our data is shown in Appendix \ref{sec:HIabs}.

\section{Sample selection and observations}
\label{sec:sample}

We use the presence of HI 21-cm absorption as a prior for there being significant column densities of (possibly cold) neutral gas in front of the radio source, which would increase our chances of detecting OH molecular absorption. 
We select our sample from the Westerbork Synthesis Radio Telescope (WSRT) HI absorption survey  \citep{maccagni17,gereb14,gereb15}, which has 64 associated HI 21-cm absorption detections in total and 14 of them have their OH 18-cm main lines (1665 and 1667 MHz) fall within the FAST frequency ranges. We observed 8 of them in this pilot project.
A target with a deep intervening absorption selected from \citet{Dutta17} is also observed for comparison. 
Parameters of the targets are listed in Table \ref{tab:fast_sample}.

\begin{table*}
	\centering
	\caption{Targets observed with FAST }
	\label{tab:fast_sample}
	\begin{tabular}{lcccrcccrrr} 
\hline
Source name & RA & Dec  & $z_{\rm opt}$ & S$_{\rm 1.4GHz}$ & log\,P$_{\rm 1.4GHz}$ & Type & $\tau_{\rm HI,peak}$ & $\int\tau_{\rm HI}$ dv  & HI FWHM  & $v_{\rm HI}$ \\
 & (deg)& (deg) & & (mJy) & (W.Hz$^{-1}$) & & & (km.s$^{-1}$) & (km.s$^{-1}$) & (km.s$^{-1}$)  \\
& (1) & (2) & (3) & (4) & (5) & (6) & (7) & (8) & (9) & (10)  \\
\hline
J084307.11+453742.8 & 130.77962 & $+$45.62856 & 0.1919 & 331.00 & 25.54 & A & 0.2730 & 25.86 & 79.2 & 57.85 \\
J112332.04+235047.8 & 170.88350 & $+$23.84661 & 0.2070 & 142.69 & 25.25 & A & 0.1048 & 5.01 & 156.2 & 219.69\\
J170815.25+211117.7 & 257.06354 & $+$21.18825 & 0.2241 & 34.36 & 24.71 & A & 0.1569 & 28.38 & 211.1 & 3.23\\	
J130132.61+463402.7 & 195.38587 & $+$46.56742 & 0.2055 & 97.00 & 25.07 & A & 0.0180 & 4.24 & 172.4 & -308.84\\
J142210.81+210554.1 & 215.54504 & $+$21.09836 & 0.1915 & 84.30 & 24.94 & A & 0.0480 & 8.36 & 179.8 & -196.05\\
J103932.12+461205.3 & 159.88383 & $+$46.20147 & 0.1861 & 30.81 & 24.48 & A & 0.0859 & 14.75 & 86.1 & 3.23\\
J153452.95+290919.8 & 233.72062 & $+$29.15550 & 0.2010 & 48.57 & 24.75 & A & 0.0275 & 6.77 & 233.7 & 26.61\\
J090325.54+162256.0 & 135.85642 & $+$16.38222 & 0.1823 & 47.81 & 24.65 & A & 0.0949 & 8.98 & 163.0 & 2.97 \\
J205449.64+004149.8$^{\rm a}$ & 313.70685 & $+$0.69718 & 0.2015 & 362.00 & 25.55 & I & 0.61 & 29.92 & 150.0 & -\\
\hline
	\end{tabular}
    \begin{flushleft}
      \small
      Columns: 
      (1) hour angle in degree; 
      (2) declination in degree; 
      (3) optical redshift from SDSS; 
      (4) 1.4 GHz continuum flux of the quasar in mJy; 
      (5) 1.4 GHz continuum flux of the quasar in physical units; 
      (6) type of the absorber, A for associated, I for intervening; 
      (7) peak HI optacity; 
      (8) integrated HI optical depth; 
      (9) full-width-half-maximum of the HI 21-cm absorption; 
      (10) HI 21-cm absorption centroid velocity from the optical redshift. \\
    Notes: 
    (a) SDSS J205449.64+004149.8 is a QSO-galaxy pair (QGP) with the QSO redshift z=0.284 \citep{Dutta17}. Other targets are selected from \citet{maccagni17}.
    \end{flushleft}
\end{table*}

FAST has finished its construction and is now taking observations \citep{Jiang20}. Its declination range is from -14.3 to +65.7 deg, and  -0.7 to +52.1 deg for  full sensitivity. The potential of FAST for extragalactic absorption study has been discussed by \citet{yu17} and \citet{li19}. FAST is currently mounted with a 19-beam receiver covering 1.05 - 1.45 GHz. The beam size is about 2.9" and the system temperature around 1.4 GHz is less than 24K for the central beam. The spectral back-end can record data in four polarizations with a frequency resolution about 7.63 kHz (65536 channels over 1.0 - 1.5 GHz). 

This paper is based on data from a `Shared-Risk' pilot observation program. The initial part of the observation for SDSS J084307.11+453742.8 was taken in normal tracking mode. We use the central beam (M01) of the 19-beam receiver to track the target for 2 minutes and then move the beam towards an OFF point for 2 minutes. The switch time between ON and OFF points is about 10 min. Later observations for other sources were taken in a newly developed ON-OFF mode. The ON-OFF mode has a much shorter ON-OFF switch time, i.e. 30 s, if the distance between ON and OFF points are within 30 arcmin. We have carefully chosen the OFF point for each source so that a side beam (M08 or M14) would be pointing to the source when the central beam is pointing to the OFF point. We do multiple ON-OFF cycles for each source and the ON-source integration time for each cycle is 2 - 3 min. The total (including central beam and side beam) integration time for each source is tabulated in Table \ref{tab:fast_results}.

We record the data every 0.1 s. For most of the observations, we inject a 10K continuum signal using a noise diode for 0.1 s every 1s. For SDSS J205449.64+004149.8, we inject the noise for 1 s every 3 min. 

\section{Data reduction and standing wave removal}
\label{sec:data}
\begin{figure*}

	\includegraphics[width=15cm]{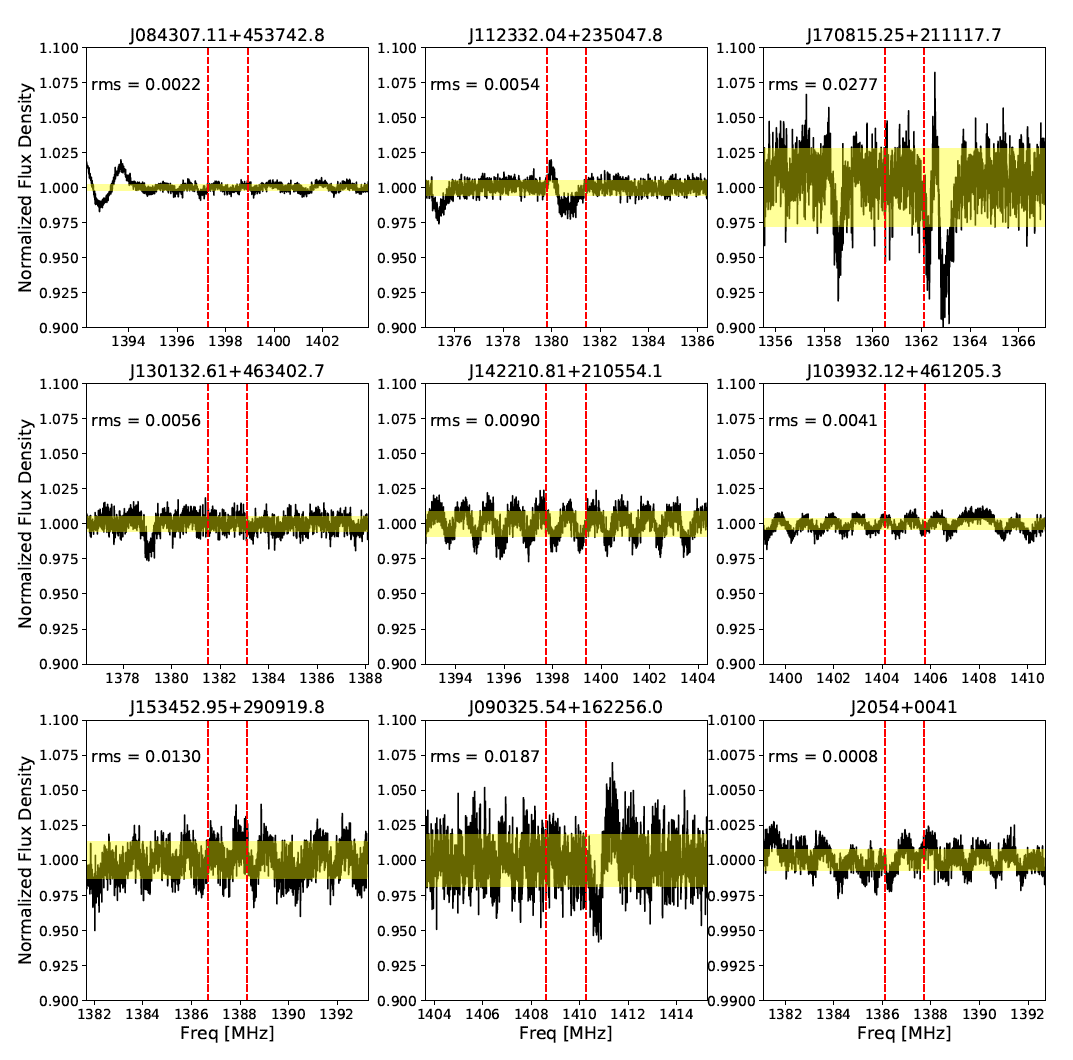}
	\caption{Initial spectra of the targets before removal of the baseline ripple. The red vertical dash lines show the expected locations of redshifted OH 1665 and 1667 lines, calculated from the optical redshifts and HI 21-cm absorption velocities. The rms values of the spectra within the shown frequency range are shown in each panel. A yellow shaded region is also overplotted in each panel to show the rms. Note the last panel (J2054+0041) has a much smaller y-axis range.}
    \label{fig:spectra_all_ripple}
\end{figure*}

We extract the noise diode signals from the data and use them to convert the spectral signals without noise diode injections into units of K. Obvious RFI is removed using a $3-\sigma$ clipping method. We then do ON-OFF subtraction and combine the first two polarizations. The results are shown in Fig. \ref{fig:spectra_all_ripple}. The spectra are normalized by dividing a straight line fitted using the continuum around the OH 1667 line vicinity. 


\begin{figure*}
	\includegraphics[width=15cm]{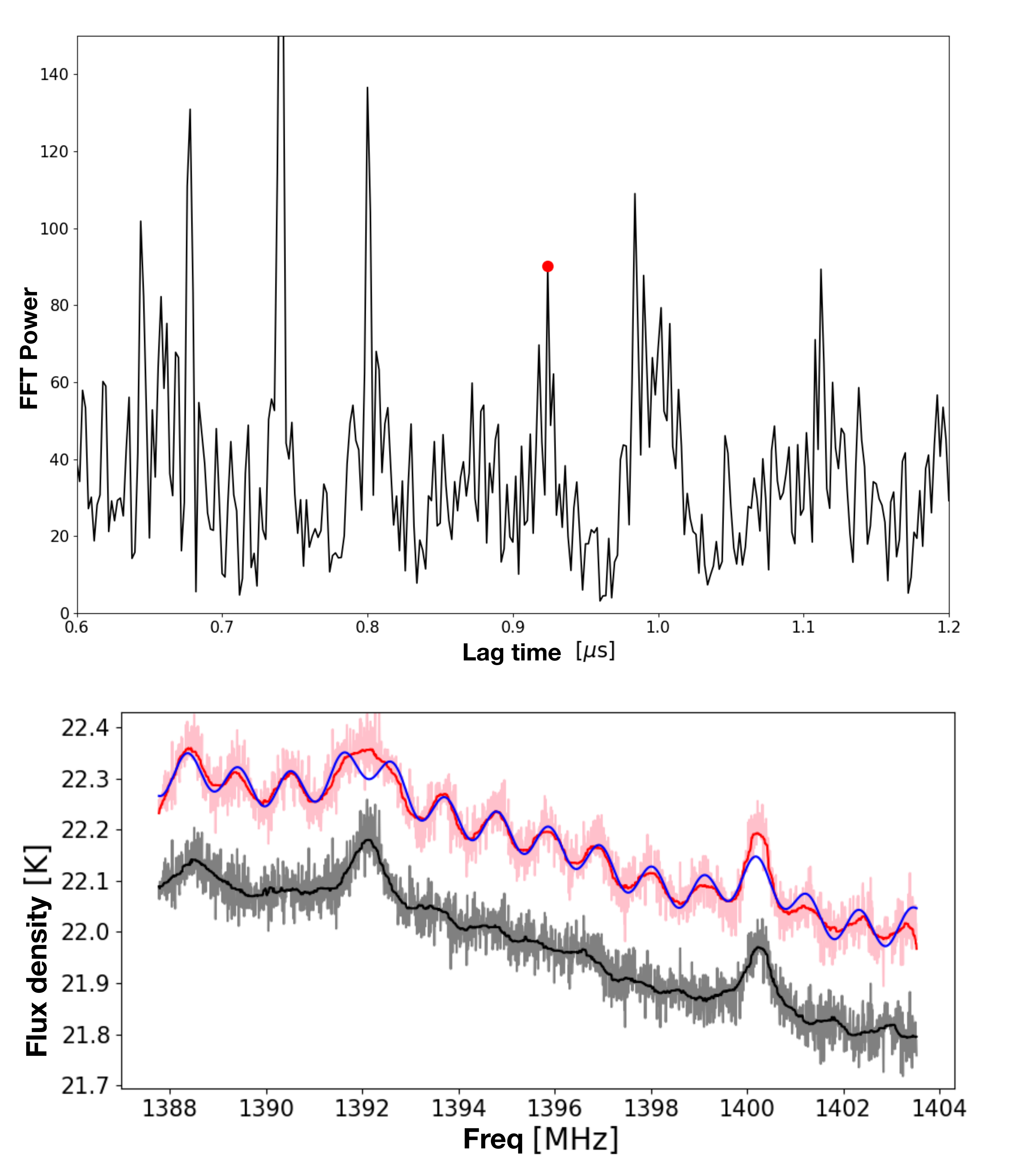}
	\caption{Example of standing wave removal. Upper panel shows the Fourier transform of a spectrum. The red dot indicates the location of the 1 MHz standing wave. Lower panel shows the spectrum before (pink) and after (gray) ripple removal. The blue line shows the sinusoidal fitting on top of the Gaussian smoothed spectrum. The red and black lines show the median filtered spectra with a filter window of 53 channel (0.4 MHz). The ripple removed spectrum is shifted down by 0.2 K in this plot for a better view.}
    \label{fig:ripple_removal}
\end{figure*}

Some of the spectra are affected by standing waves with a frequency of about 1 MHz, which corresponds to a velocity width of about 200 km/s at 1600 MHz. This 1 MHz standing wave could be introduced by reflections of leaked signals between the receiver and the bottom panel of the telescope. Improvements in hardware have been made to reduce the reflections but the standing wave still exists. We thus remove these 1 MHz ripples using the following prescription. 
(1) Shape of the ripple is relatively stable in spectra taken within one observation cycle (2-3 mins). So we do Fourier transform on the original spectrum from each observation cycle and identify the peak location $t_p$ within 0.91 and 0.95 $\mu s$. 
(2) We smooth the spectrum using a Gaussian kernel and subtract the smoothed spectrum from the original spectrum. The width (standard deviation) of the Gaussian kernel is chosen to be 60 channels (about 0.46 MHz) so that the smoothed curve is free of the 1 MHz ripple but still retains features with widths larger than 1 MHz. 
(3) We fit the resulting spectrum (excluding the 1 MHz region around the OH 1667 MHz line) using a sinusoidal function with a fixed period of $1/t_p$. 
(4) We then obtain the ripple removed spectrum by subtracting the fitted sinusoidal function from the original spectrum. 
An example of this fitting process is shown in Fig.\ref{fig:ripple_removal}.

\section{Results and discussions}
\label{sec:results}
\begin{figure*}
	\includegraphics[width=15cm]{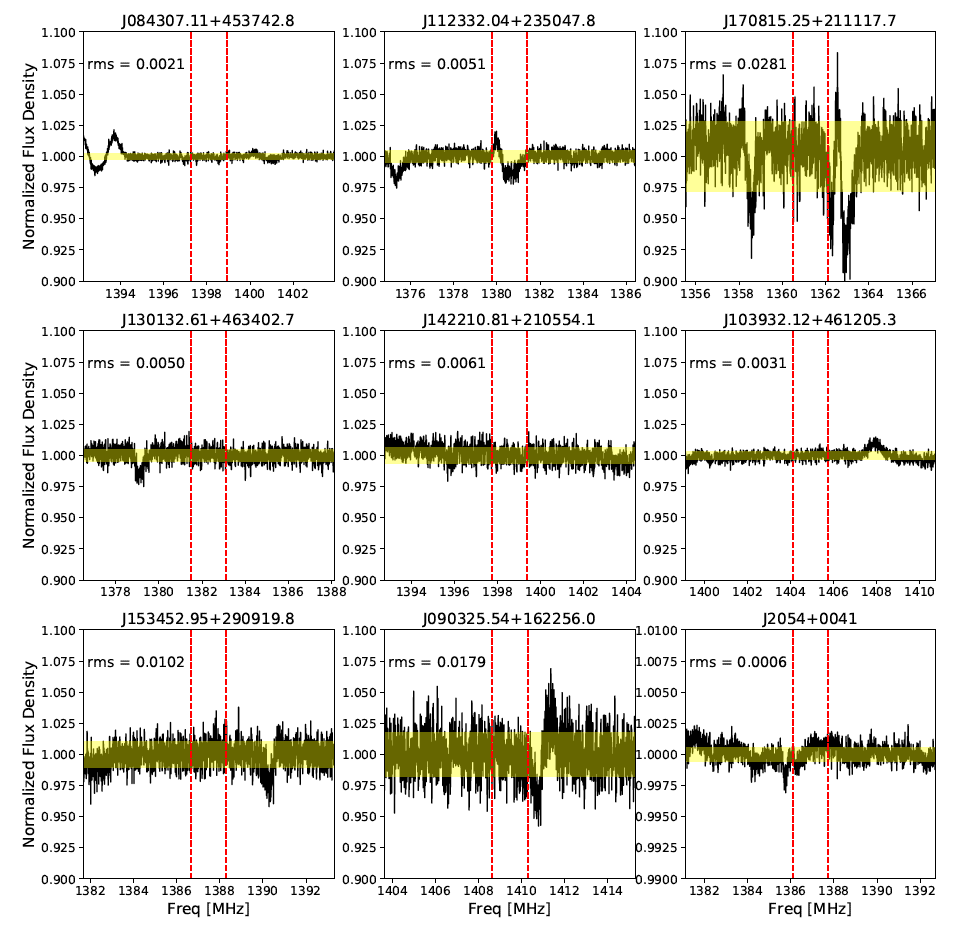}
	\caption{
	Final spectra, as in Fig. \ref{fig:spectra_all_ripple} but with the baseline ripple removed.
	Note that the dips in these panels are all caused by RFI. }
    \label{fig:spectra_all_noripple}
\end{figure*}

The final reduced spectra with standing waves removed are shown in Fig. \ref{fig:spectra_all_noripple}. The locations of the redshifted OH 1665 and 1667 lines are marked by red vertical dash lines. Note that the dips around the vertical lines can be attributed to RFI.
We do not see any robust OH absorption signal. The rms values around the OH 1667 line are also shown in each panel. Five sources  are affected by RFI, while the remaining four are relatively clean.
The derived 3$-\sigma$ integrated optical depth upper-limits range from 0.03 to 1.6, assuming the HI line width. The extremely low limit of 0.03 resulted from a two hour integration toward J205449.64+004149.8.
Observation details and rms values are also tabulated in Table \ref{tab:fast_results}.

\begin{table*}
	\centering
	\caption{optical depth upper limits of FAST spectra }
	\label{tab:fast_results}
	\begin{tabular}{lcccc} 
\hline

Source name & $t_{\rm int}$  &  $\tau_{1667,\rm rms}$ & Velocity resolution   & $3 \int\tau_{\rm 1667,rms}$ dv  \\

  & (min) & ($10^{-2}$)  & (km\,s$^{-1}$) & (km\,s$^{-1}$) \\
 
  & (1) & (2) & (3) & (4) \\
\hline
J084307.11+453742.8 & 30  & 0.21 & 1.64 & 0.07  \\
J112332.04+235047.8 & 19  & 0.51 & 1.66 & 0.25  \\
J170815.25+211117.7 & 24  & 2.81 & 1.68 & 1.59 \\
J130132.61+463402.7 & 24  & 0.50 & 1.65 & 0.25  \\
J142210.81+210554.1 & 24  & 0.61 & 1.64 & 0.31  \\
J103932.12+461205.3 & 22  & 0.31 & 1.63 & 0.11 \\
J153452.95+290919.8 & 23  & 1.02 & 1.65 & 0.60  \\
J090325.54+162256.0 & 16  & 1.79 & 1.62 & 0.87  \\
J205449.64+004149.8 & 132 & 0.06 & 1.65 & 0.03 \\
\hline \\
	\end{tabular}
    \begin{flushleft}
      \small

      Columns:
      (1) On-target integration time. 
      (2) rms of the normalized spectra after standing wave removal around the OH 1667 line (within 1 MHz). 
      (3) Velocity resolution around the observed frequency of OH 1667 line. 
      (4) $3\sigma$ upper limit of integrated OH 1667 optical depth within HI FWHM. 
    \end{flushleft}
\end{table*}

\begin{figure*}
	\includegraphics[width=7.5cm]{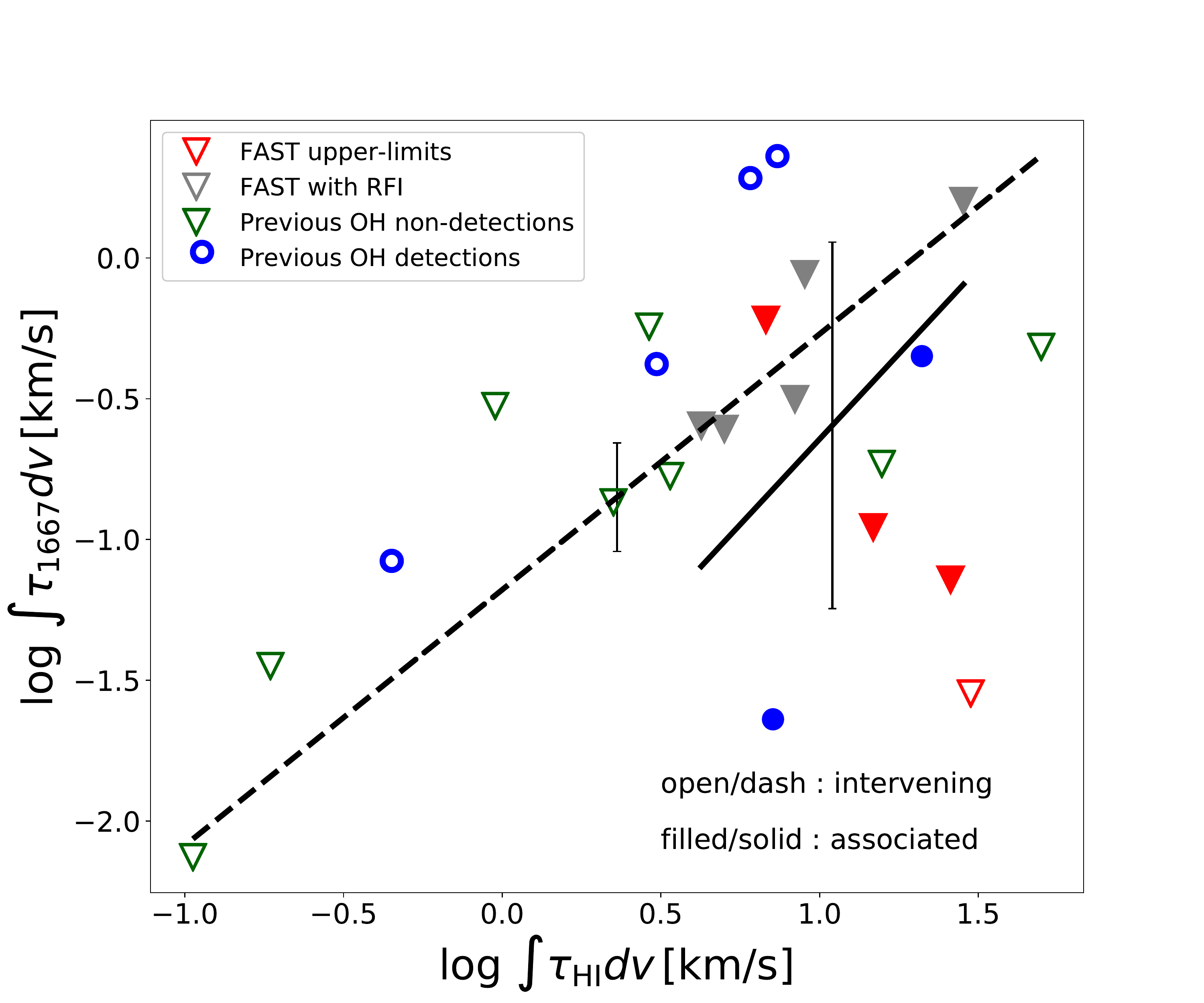}
	\includegraphics[width=7.5cm]{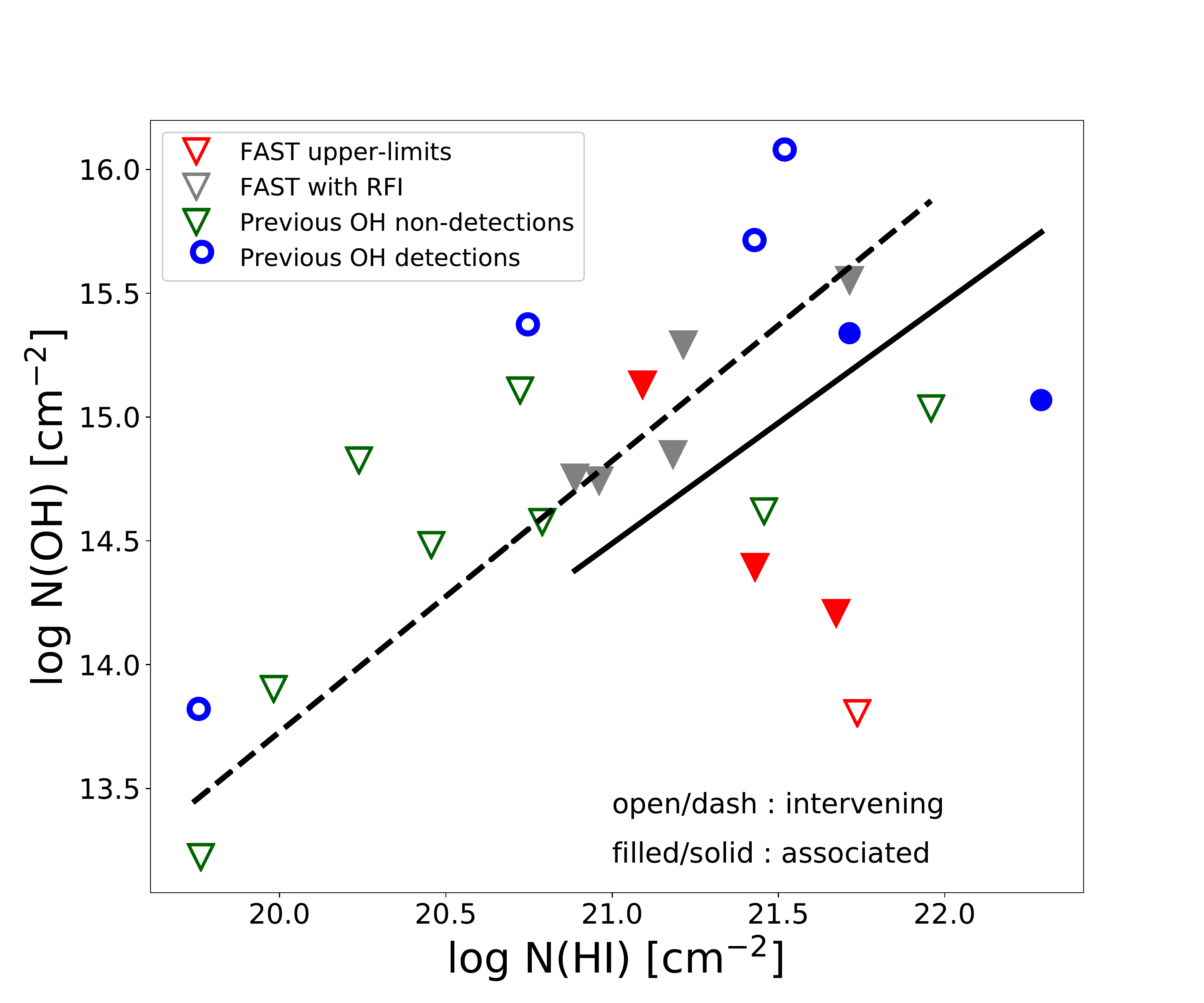}
	\caption{Left panel: Integrated HI optical depth versus integrated OH 1667 optical depth; right panel: HI column density versus OH column density. Red triangles are 3-$\sigma$ upper limits derived using newly observed FAST data, gray triangles are also upper limits from FAST data with obvious RFI, green triangles are upper limits calculated using the spectral rms and HI FWHM provided by \citet{Gupta18} and references therein, blue dots are the 6 OH absorption detections listed in Table \ref{tab:archival_OH}. Open symbols are intervening absorbers and filled symbols are associated absorbers. The black solid and dash lines show a linear fit to the associated and intervening absorbers respectively using a survival statistics \citep{Feigelson85} with the Schmitt binning method \citep{Schmitt85}. 
	The 1-$\sigma$ uncertainties for the fittings 
	on the y-axis are shown as error bars. We do not
	show the error bars for the right panel because the uncertainties, 5.61 and 6.61 for the solid and dash lines respectively, are much larger than the y-axis range shown here.}
    \label{fig:OH_HI_int}
\end{figure*}

We plot the integrated HI optical depth versus integrated OH 1667 optical depth in the left panel of Fig. \ref{fig:OH_HI_int}, together with detections and upper limits from \citet{Gupta18}.  
We show linear fits to both associated and intervening absorbers using a survival statistics \citep{Feigelson85} with the Schmitt binning method \citep{Schmitt85}, which can deal with detections and upper limits simultaneously. 
The fitted relation for the associated absorbers is 
$\log \int{\tau_{1667}}dv = (-1.86\pm0.64) + (1.21\pm0.61)\times \log \int{\tau_{\rm HI}}dv$, with the generalized Spearman's correlation parameter $\rho = 0.17$ and a probability of $60\%$ for a null correlation.
The fitted relation for the intervening absorbers is 
$\log \int{\tau_{1667}}dv = (-1.18\pm0.20) + (0.91\pm0.30)\times \log \int{\tau_{\rm HI}}dv$, with $\rho = 0.28$ and a probability of $33\%$ for a null correlation.
Both associated and intervening absorbers show a weak positive correlation between integrated HI and OH optical depths from the fitting, however, 
the fitting uncertainties are large and the statistics
may not be reliable because of small sample numbers,
especially for the associated absorbers.
The associated absorbers generally have lower OH optical depths than intervening absorbers with similar HI optical depths: the mean integrated OH 1667 optical depth of the detected intervening and associated absorbers are 1.13 and 0.24 respectively.

We calculate the HI and OH column densities using
\begin{equation}
    {\rm N(HI)} = 1.823 \times 10^{18} \, \frac{ T^{\rm HI}_x}{f^{\rm HI}_c} \int \tau_{\rm HI} dv \,\,cm^{-2},
\end{equation}
and 
\begin{equation}
    {\rm N(OH)} = 2.24 \times 10^{14} \, \frac{ T^{\rm OH}_x}{f^{\rm OH}_c} \int \tau_{1667} dv \,\,cm^{-2},
\end{equation}
where $T^{\rm HI}_x$ and $T^{\rm OH}_x$ are excitation temperatures and $f^{\rm HI}_c$ and $f^{\rm OH}_c$ covering factors of HI and OH. We use $T^{\rm HI}_x = 100 {\rm K}$, $T^{\rm OH}_x = 10 {\rm K}$, $f^{\rm HI}_c = 1$ and $f^{\rm OH}_c = 1$ for all our targets, which are generally used in most previously detected absorbers at cosmological distances \citep[e.g.][]{KC02,Grasha19}.
We plot the column densities in the right panel of Fig. \ref{fig:OH_HI_int}.
 \citet{Gupta18} claims that their OH column densities are much lower than previously detected OH absorbers but we see no gap between the upper limits and detections.
Note they assume the OH velocity widths are 2 km/s and the OH excitation temperature $T^{\rm OH}_x = 3.5 {\rm K}$. Here we assume the velocity widths of the OH 1667 lines are the same as their HI 21-cm absorption lines and $T^{\rm OH}_x = 10 {\rm K}$, $f_c^{\rm OH} = 1$.  
We also perform survival statistical analysis to the HI and OH column densities. 
The fitted relation for the associated absorbers is 
$\log {\rm N(OH)} = (-5.97\pm 5.61) + (0.97\pm0.27)\times \log {\rm N(HI)}$, with the generalized Spearman's correlation parameter $\rho = 0.68$ and a probability of $4\%$ for a null correlation.
The fitted relation for the intervening absorbers is 
$\log \rm{N(OH)} = (-8.17\pm6.16) + (1.09\pm0.30)\times \log {\rm N (HI)}$, with $\rho = 0.12$ and a probability of $67\%$ for a null correlation.
They are consistent with the relationship between the integrated HI and OH optical depths.

\begin{figure*}
	\includegraphics[width=7.5cm]{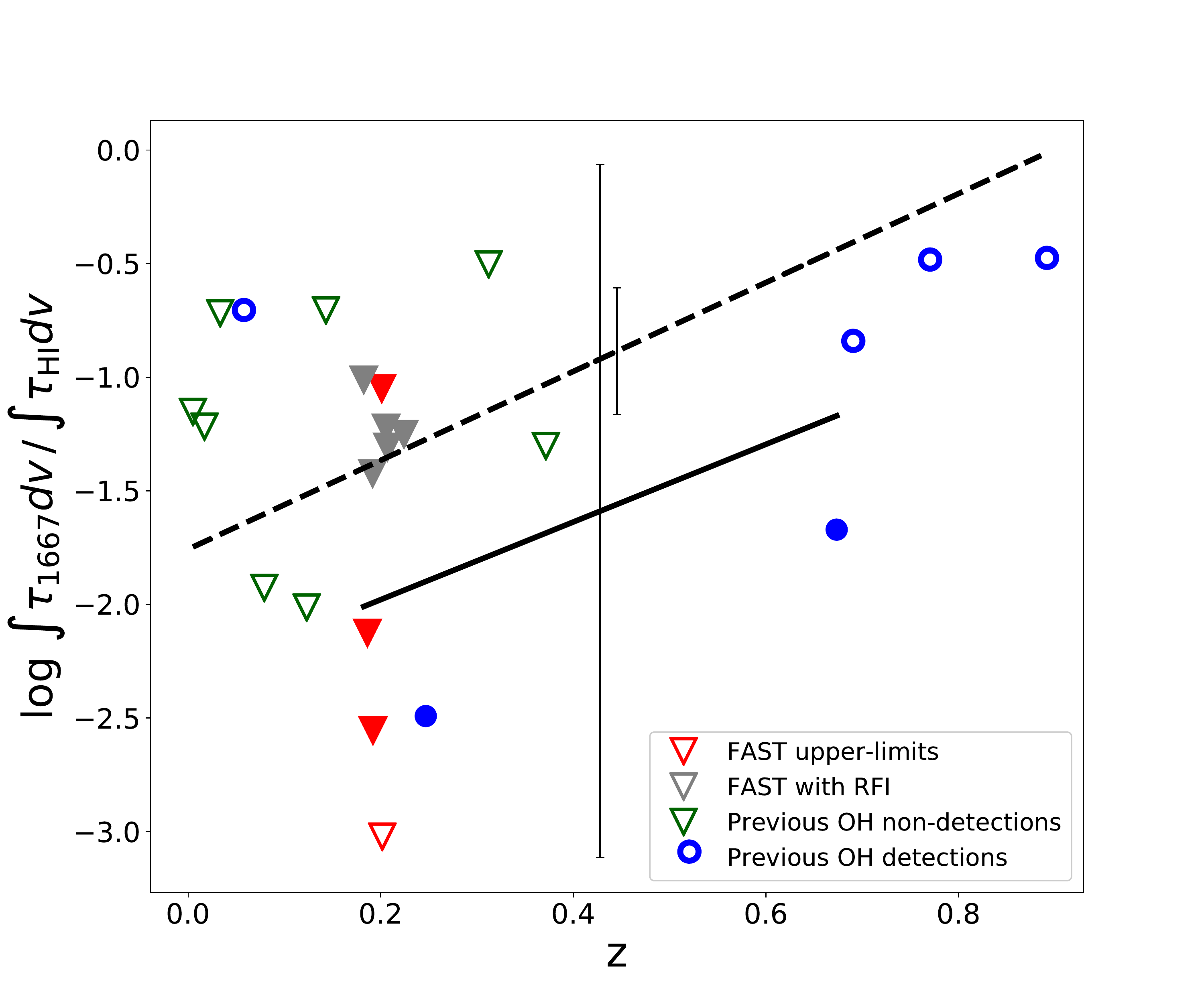}
	\includegraphics[width=7.5cm]{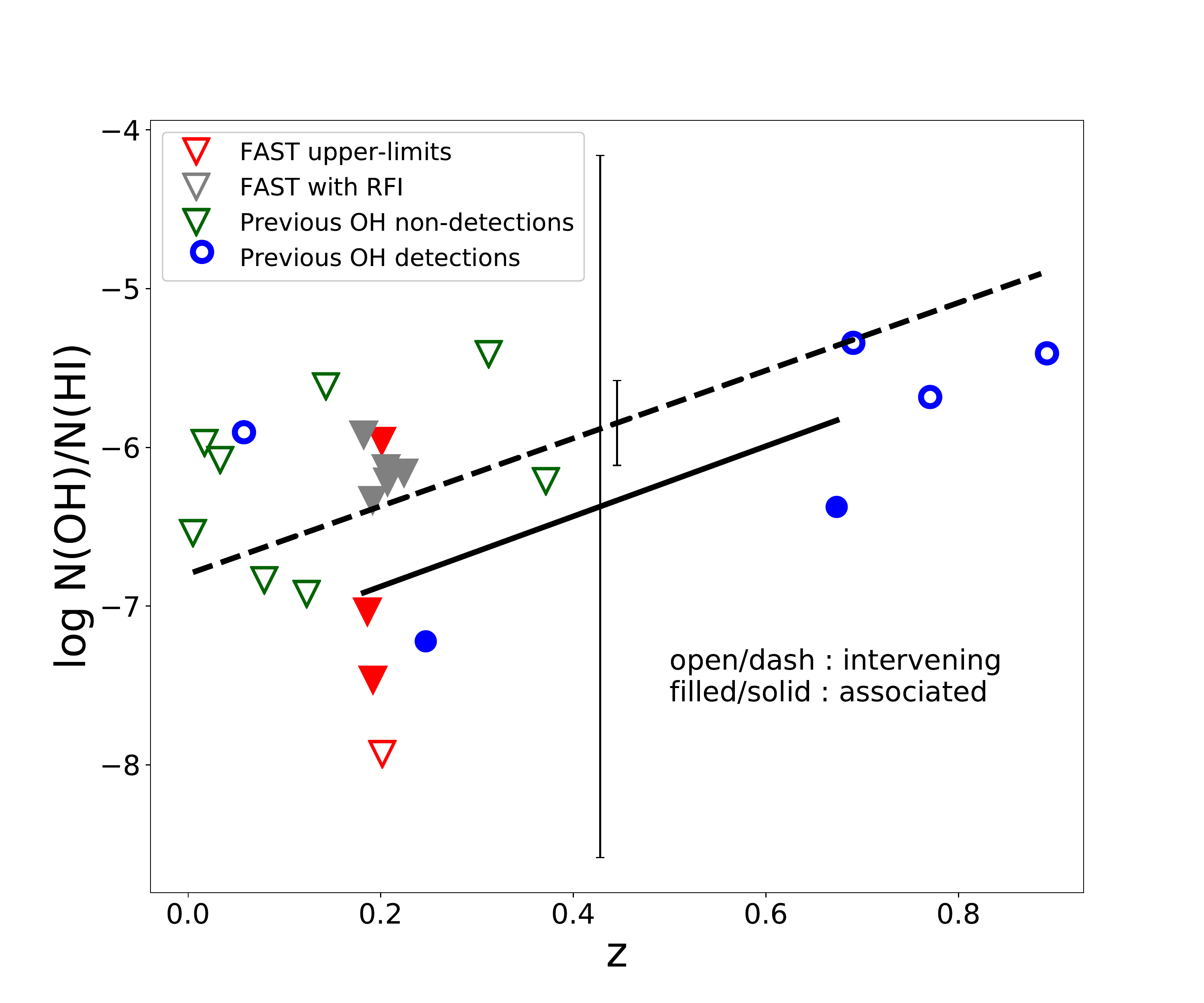}

	\caption{Integrated OH and HI optical depth ratio (left panel) and OH/HI column density ratio (right panel) versus redshift. Symbols are the same as Fig. \ref{fig:OH_HI_int}. }
    \label{fig:nratio_vs_z}
\end{figure*}

We explore the evolution of OH abundance by plotting the OH/HI optical depth and column density ratios versus redshift in Fig. \ref{fig:nratio_vs_z}.
Although we show in Fig. \ref{fig:OH_HI_int} that the gap between OH column densities calculated by \citet{Gupta18} and previous high redshift detections disappeared if using a uniform $T_x^{\rm OH}$ and a wider velocity width, Fig. \ref{fig:nratio_vs_z} show that absorbers with a redshift less than 0.4 still have a lower OH/HI column density ratio than those with a higher redshift. 
The linear fitting using the survival statistics show that both associated and intervening absorbers have a increasing OH abundance with increasing redshift. 
The fitted relations for the associated absorbers are
$\log \int{\tau_{1667}}dv / \int{\tau_{\rm HI}}dv = (-2.32\pm1.66) + (1.71\pm7.79) \times z $,  
and 
$\log {\rm N(OH)/N(HI)} = (-7.32\pm1.39) + (2.21\pm6.36) \times z$;
and the generalized Spearman's $\rho = $ 0.57 and 0.56, and a null correlation probability of $9\%$ and $5\%$ respectively.
The fitted relations for the intervening absorbers are 
$\log \int{\tau_{1667}}dv / \int{\tau_{\rm HI}}dv = (-1.76\pm0.29) + (1.96\pm0.62)\, z $,  
and 
$\log {\rm N(OH)/N(HI)} = (-6.80\pm0.24) + (2.14\pm0.43)\, z$;
with $\rho = $ 0.57 and 0.65, and a null correlation probability of $9\%$ and $2\%$ respectively.
This implies that the molecular fraction has decreased from higher redshifts down to $z<0.4$ and this could be the reason of the star formation rate decline since $z=2$ \citep{MD14}.  
It is consistent with previous studies such as
\citet[][and references therein]{Tacconi20}, which show that the CO traced molecular gas content decreased since redshift $\sim 1.4$. However, we note that OH is a tracer of diffuse molecular gas, as opposed to CO being a more direct tracer of dense molecular clouds and star formation \citep[e.g.][]{vanDishoeck86,allen15,Xu16}.

Figs. \ref{fig:OH_HI_int}-\ref{fig:nratio_vs_z} show that most associated absorbers have a lower OH abundance comparing to intervening absorbers. \citet{Curran16} claims that associated absorbers are mostly originated from high-velocity nuclear gas feeding the central AGN. This implies that the AGNs in these sources are not promoting OH molecule formation if not considering the effects of covering factors and excitation temperatures. 

We note that the correlations derived above using the survival statistics have large uncertainties. 
Furthermore, the reported generalized Spearman's correlation parameters may not be reliable because the data points used in the analysis are less than 30.  We expect a further observation of at least more than 20 absorbers spanning a wider redshift range in the future in hope to reveal a more reliable relationship between OH abundance and redshifts. This could potentially provide a better constrain to the molecular
abundance evolution in galaxies.

\begin{figure*}

	\includegraphics[width=7.5cm]{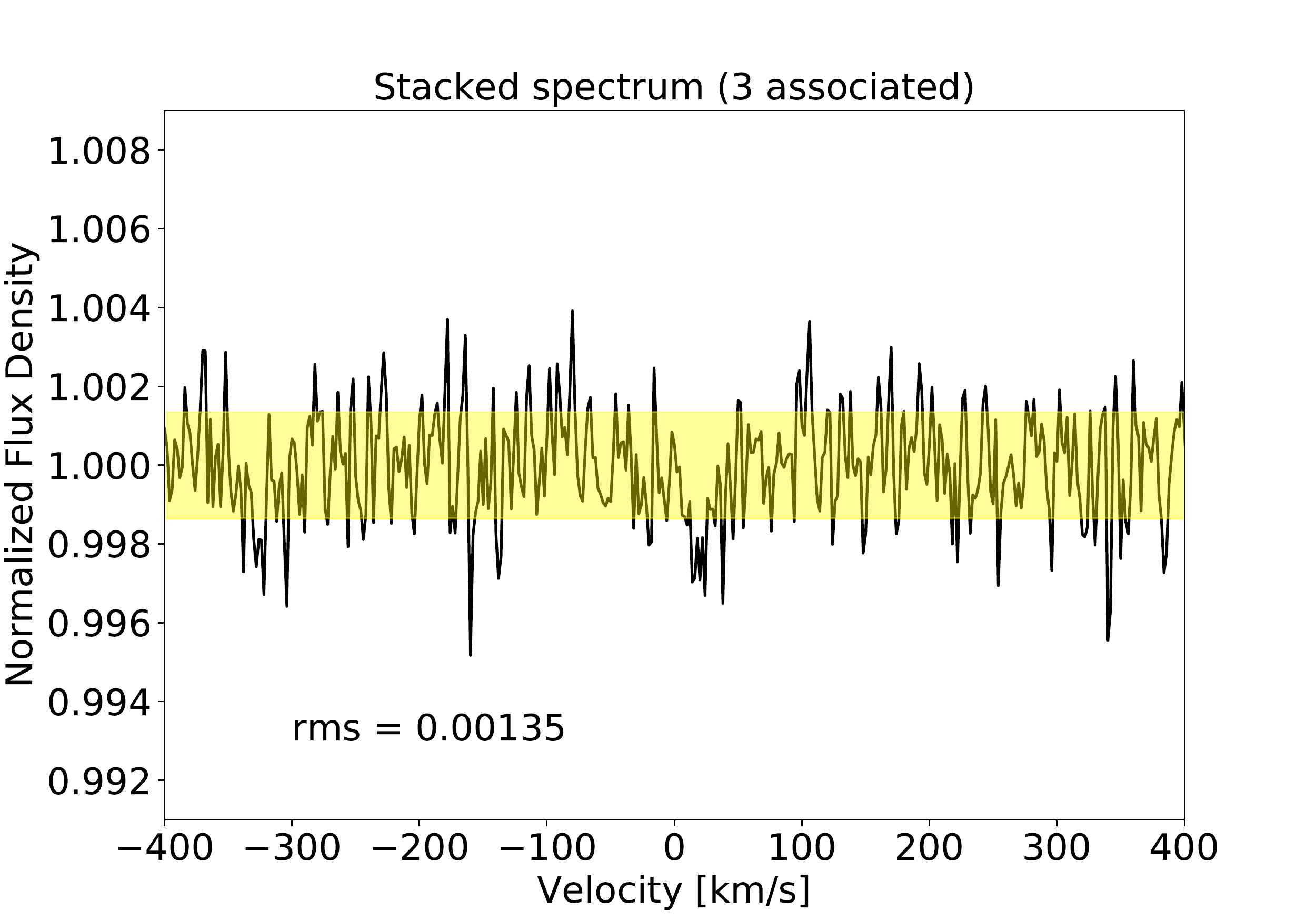}
	\includegraphics[width=7.5cm]{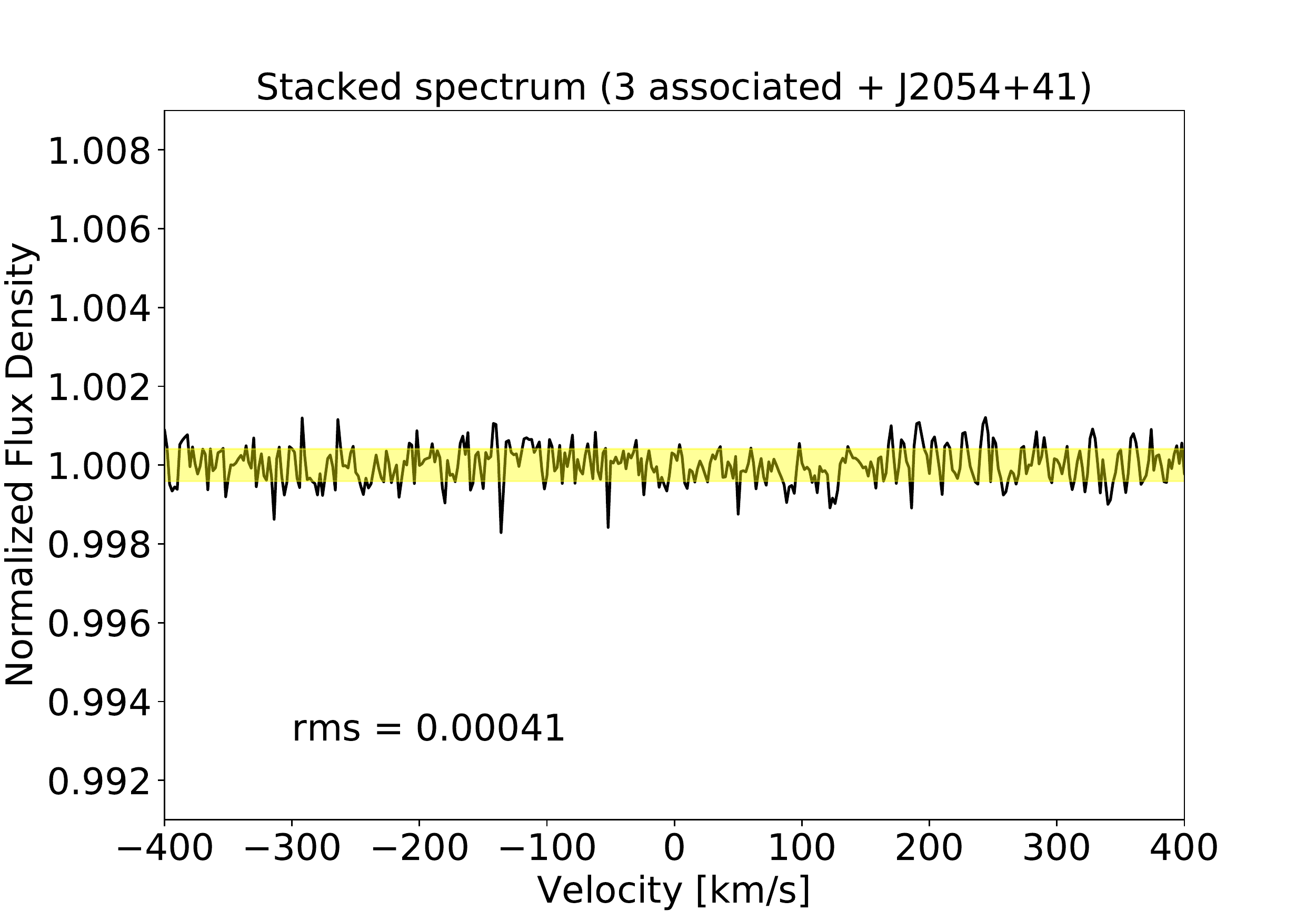}
	\caption{Stacked spectra. Left panel shows the stacked spectrum of the three associated absorbers J084307.11+453742.8, J153452.95+290919.8, J103932.12+461205.3, which are free of RFI at the vincinity of their OH 1667 lines. Right panel shows the stacked spectrum of the three plus J205449.64+004149.8. The spectra are centered at the OH 1667 line and has a velocity resolution of 2 km/s. A yellow shaded region is overplotted to show the rms.}
    \label{fig:stack}
\end{figure*}

We stack the ripple removed spectra of 3 of our associated absorbers which are free of RFI, i.e. J084307.11+453742.8, J153452.95+290919.8, J103932.12+461205.3. We weight each individual spectrum using their inverse variance ($1/rms^2$) during stacking. The normalized stacked spectrum is shown in the left panel of Fig. \ref{fig:stack}. There is a dip in the stacked spectrum around 0 km/s but the amplitude is well within 3-$\sigma$ of the spectrum. The 3-$\sigma$ OH column density upper limit derived using the stacked spectrum rms = 0.00135, with an assumed OH velocity width of 150 km/s, $\rm T_x^{\rm OH} = 10 K$ and $f_c^{\rm OH} = 1$, is about $1.57\times 10^{14} \rm cm^{-2}$, which are consistent with previous values \citep{KC02, Kanekar03, Kanekar05, Gupta18}. Using a mean HI column density of the four targets, we obtain an OH abundance upper limit [OH]/[HI] < $\sim 5.45\times 10^{-8}$ for these associated absorbers, which is only 50\% of the typical Galactic value \citep{nguyen18} and represents the deepest extragalactic limit to the OH abundance achieved by OH absorption measurement at cosmological distances.
We further added the spectrum of the intervening absorber J205449.64+004149.8 in the stacking and the result is shown in the right panel of Fig. \ref{fig:stack}. The rms of the stacked spectrum, 0.00041, is then dominated by J205449.64+004149.8. We do not see any obvious OH 1667 absorption line in the stacked spectrum either.

\section{Summary}
\label{sec:summary}
We derive upper limits of OH column densities for 8 associated and 1 intervening absorbers at redshifts z$\in$ [0.1919,0.2241] through absorption measurements with FAST. The combined upper limits to the OH abundance for 3 associated absorbers, [OH]/[HI] < $\sim 5.45\times 10^{-8}$, is the lowest ever achieved. The associated absorbers have a slightly lower OH abundance than their intervening counterparts. The results are largely consistent with a trend of decreasing OH abundance with decreasing z.

\section*{Acknowledgements}
We thank Carl Heiles and Marko Krco for useful discussions. We also thank the anonymous referee for very helpful comments and extremely efficient responses.  
This work is supported by the National Natural Science Foundation of China grant No. 11988101, No. 11725313, U1931110, 11703036, the International Partnership Program of Chinese Academy of Sciences grant No. 114A11KYSB20160008, and the CAS Interdisciplinary Innovation Team(JCTD- 2019-05). 
Parts of this research were supported by the Australian Research Council Centre of Excellence for All Sky Astrophysics in 3 Dimensions (ASTRO 3D), through project  number CE170100013. 
ZZ also acknowledges support from an ACAMAR visiting fellowship.

\section*{Data Availability}
The data underlying this article are available in the project 2019a-035-S at https://fast.bao.ac.cn/cms/article/95/, and can be shared on request to the FAST data center or to the corresponding author.



\bibliographystyle{mnras}
\bibliography{FAST_OH_paper} 





\appendix

\section{HI 21-cm absorption of J084307.11+453742.8}
\label{sec:HIabs}
\begin{figure*}

	\includegraphics[width=13cm]{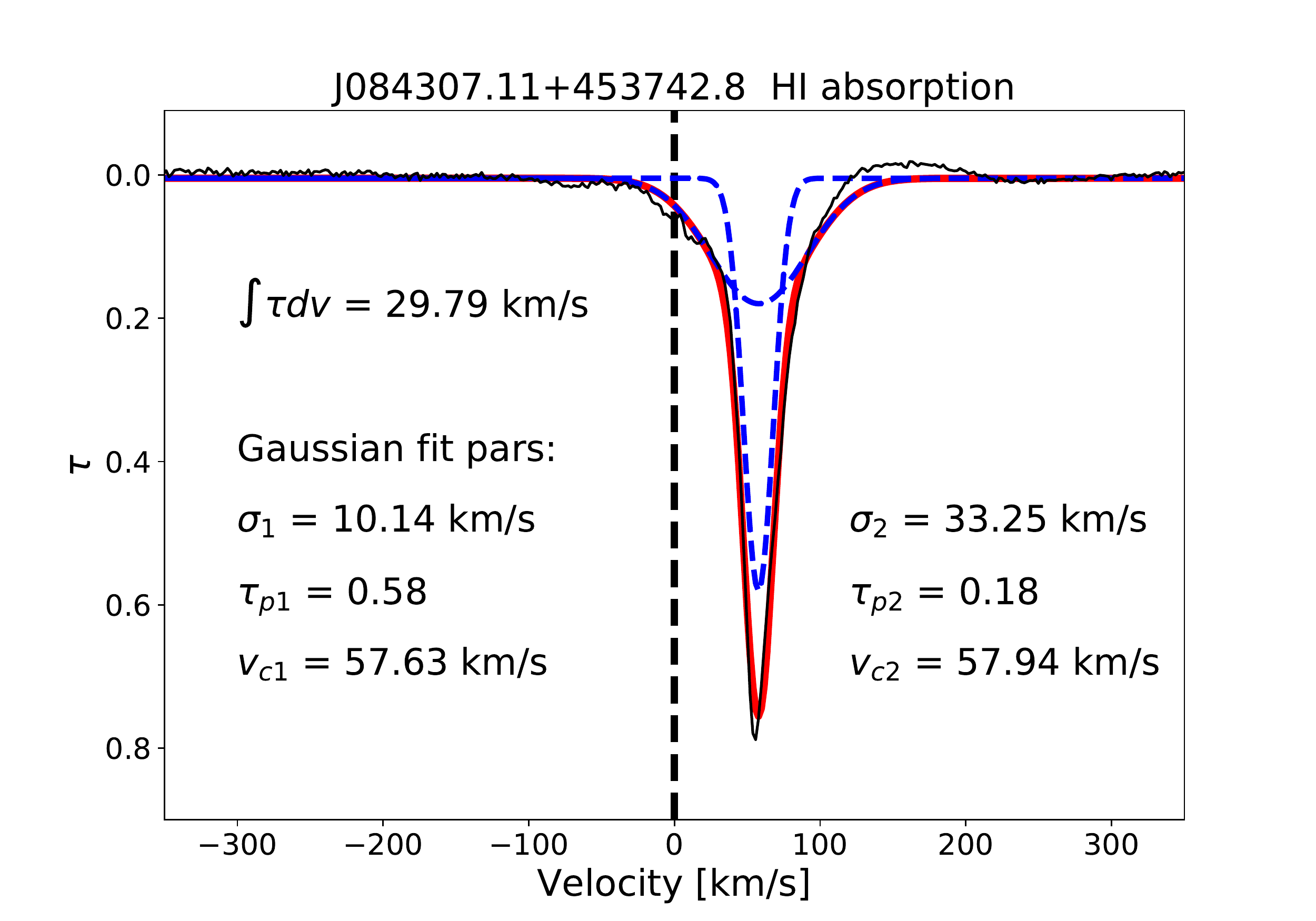}
	\caption{FAST HI 21-cm absorption spectrum of J084307.11+453742.8. The blue dash lines show the two fitted  Gaussian components, the red line shows the combined fit and the vertical dash line shows the optical velocity of the target.
	}
    \label{fig:J0843HI}
\end{figure*}

Although most of the HI frequencies of our targets locate within the 1150 - 1250 MHz frequency range, which is severally affected by satellite RFI, one of our target, J084307.11+453742.8, does show a deep HI 21-cm absorption. The HI absorption feature is shown in Fig.\ref{fig:J0843HI}. The integrated optical depth from -200 km/s to 200 km/s is about 29.79 km/s, which is a little bit larger than the value from \citet{maccagni17} (Table \ref{tab:fast_sample}). 
We fit the absorption profile with two Gaussian components and the resulting peak optical depths and FWHMs are $\tau_{p1} = 0.58$, FWHM$_1 = 23.83$ km/s, $\tau_{p2} = 0.18$, FWHM$_2 = 78.13$ km/s.  The narrow component has a much deeper and narrower absorption profile comparing to \citet{maccagni17}. A possible explanation for the difference could be that an extra compact component with a proper motion moving into the line of sight. However, it could also be due to measurement error or scattering of the absorbing region due to the ISM in either the host galaxy or the Milky Way.

\bsp	
\label{lastpage}
\end{document}